\documentclass[rsi, amsmath,amssymb, reprint]{revtex4-1}

\usepackage{graphicx}
\usepackage{dcolumn}
\usepackage{bm}
\usepackage{threeparttable}
\usepackage{multirow}
\usepackage{color}
\usepackage{pifont}
\usepackage{dcolumn}
\usepackage[utf8]{inputenc}
\usepackage[T1]{fontenc}
\usepackage{mathptmx}
\usepackage{color}
\usepackage{diagbox}
\usepackage[]{lineno} 
\usepackage{float}

\begin{document}

\title{A newly-designed femtosecond KBe$_2$BO$_3$F$_2$ device with pulse duration down to 55 fs  for  time- and angle-resolved photoemission spectroscopy}

\author{Haoyuan Zhong}
\email{These authors contributed equally to this work}
\affiliation{State Key Laboratory of Low-Dimensional Quantum Physics and Department of Physics, Tsinghua University, Beijing 100084, People’s Republic of China}

\author{Changhua Bao}
\email{These authors contributed equally to this work}
\affiliation{State Key Laboratory of Low-Dimensional Quantum Physics and Department of Physics, Tsinghua University, Beijing 100084, People’s Republic of China}
\author{Tianyun Lin}
\affiliation{State Key Laboratory of Low-Dimensional Quantum Physics and Department of Physics, Tsinghua University, Beijing 100084, People’s Republic of China}
\author{Shaohua Zhou}
\affiliation{State Key Laboratory of Low-Dimensional Quantum Physics and Department of Physics, Tsinghua University, Beijing 100084, People’s Republic of China}
\author{Shuyun Zhou}
 \email{syzhou@mail.tsinghua.edu.cn}
 \affiliation{State Key Laboratory of Low-Dimensional Quantum Physics and Department of Physics, Tsinghua University, Beijing 100084, People’s Republic of China}
\affiliation{Frontier Science Center for Quantum Information, Beijing 100084, People’s Republic of China}

\date{\today}

\begin{abstract}
Developing a widely tunable vacuum ultraviolet (VUV) source with sub-100 femtoseconds (fs) pulse duration is critical for ultrafast pump-probe techniques such as time- and angle-resolved photoemission spectroscopy (TrARPES). While a tunable probe source with photon energy of 5.3--7.0 eV has been recently implemented for TrARPES by using a KBe$_2$BO$_3$F$_2$ (KBBF) device, the time resolution of 280--320 fs is still not ideal, which is mainly limited by the duration of the VUV probe pulse generated by the KBBF device.  Here, by designing a new KBBF device which is specially optimized for fs applications, an optimum pulse duration of 55 fs is obtained after systematic diagnostics and optimization. More importantly, a high time resolution of 81--95 fs is achieved for TrARPES measurements covering the probe photon energy range of 5.3--7.0 eV, making it particularly useful for investigating the ultrafast dynamics of quantum materials. Our work extends the application of KBBF device to ultrafast pump-probe techniques with the advantages of both widely tunable VUV source and ultimate time resolution.

\end{abstract}

\maketitle

\section{\label{sec:level1}INTRODUCTION}

The generation of a tunable vacuum ultraviolet (VUV) source  by a KBe$_2$BO$_3$F$_2$ (KBBF) device \cite{kbbf1, chen2009deep}
 with a widely tunable photon energy up to 7.0 eV has provided new opportunities for spectroscopic and microscopic applications that utilize VUV sources, such as angle-resolved photoemission spectroscopy (ARPES) \cite{kbbf_shin,kbbf_xj,kbbf_ak,kbbf_re1,kbbf_re2}, photoemission electron microscopy (PEEM) \cite{KBBF_PEEM1}, Raman spectroscopy and photoluminescence spectroscopy \cite{peng2018duv}.  Over the past decades, VUV sources generated by KBBF devices with prism-coupled technique have significantly advanced laser-based ARPES with a higher photon energy  and a better energy resolution than what is achieved by conventional solid-state nonlinear crystals \cite{kbbf_re1}. However, due to the large KBBF crystal thickness of $\sim$ 1 mm and the complex device structure with large coupling prisms  \cite{chen2002second},  the VUV pulses generated by KBBF device typically have a pulse duration approaching picosecond (ps) even when using a femtosecond (fs) pulse laser \cite{yang2019time,yang2013pulse}. Such significantly broadened VUV pulse has restricted its applications in  ultrafast dynamic studies via pump-probe techniques such as time- and angle-resolved photoemission spectroscopy (TrARPES).

TrARPES is a powerful technique for capturing the electronic dynamics of quantum materials with unique energy-, momentum- and time-resolved information. The implementation of a  VUV probe source with widely tunable photon energy is important for TrARPES for at least two reasons. Firstly, the tunable probe photon energy allows to measure the electronic structure at different out-of-plane momentum (k$_z$) values, since in ARPES measurements, each single probe photon energy corresponds to only one k$_z$ value \cite{zxrev2003,zxrev,ARPESPrimers}. Such k$_z$-resolving capability is particularly important for three-dimensional (3D) quantum materials such as 3D Dirac or Weyl semimetals, where the Dirac or Weyl nodes exist only at specific k$_z$ values  \cite{NB2014SCI,NB2015SCI,DingHRMP2021}. Secondly, even for two-dimensional (2D) and one-dimensional (1D) materials with relatively weak k$_z$ dispersion such as layered topological materials WTe$_2$  \cite{kzwte2}, ZrTe$_5$ \cite{kzzrte5} and 1D material TaNiTe$_5$  \cite{kztnt}, the tunable probe photon energy is also useful because it provides opportunities for accessing the full set of electronic structures, which may be modulated or suppressed by the dipole matrix element \cite{zxrev2003,zxrev,ARPESPrimers}. Therefore, a tunable probe photon energy with an ultrashort pulse duration is highly desired for TrARPES.

\begin{figure*}[htbp]
\centering
\includegraphics[] {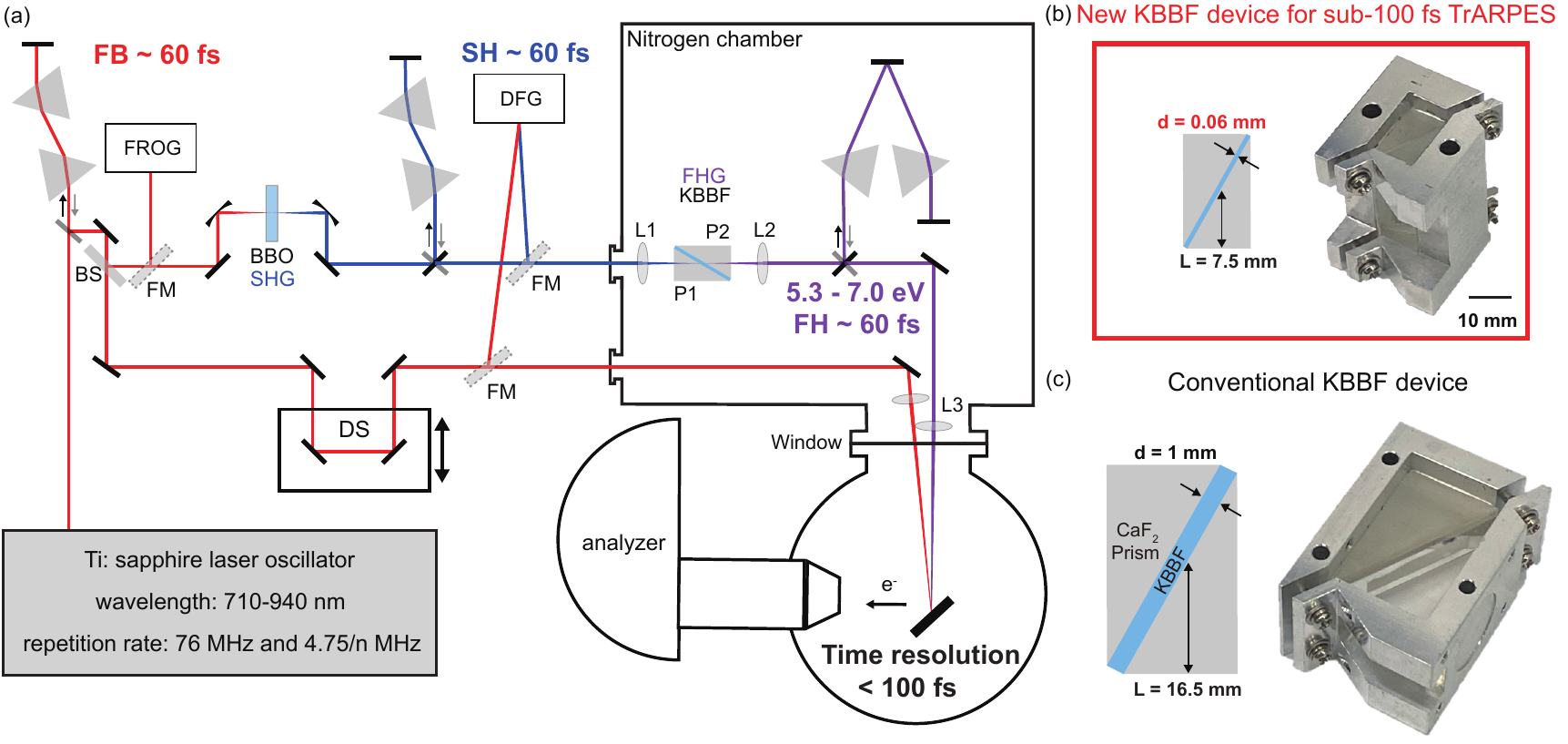}
\label{Fig1}
\caption{Schematic of the TrARPES setup with newly designed KBBF device. (a) Layout of TrARPES system. FROG: frequency-resolved optical gating;  DFG: difference frequency generation; BS: beam splitter; FM: flip mirror; DS: delay stage; L1--L3: Lenses 1--3; P1, P2: prism 1,2; SHG: second harmonic generation; FHG: fourth harmonic generation. (b) A schematic drawing and a photo of new KBBF device for fs TrARPES with thin KBBF crystal (thickness d = 0.06 mm) and small coupling prisms. (c) A schematic drawing and a photo of typical KBBF device with thick KBBF crystal (thickness d = 1 mm) and large coupling prisms.}
\end{figure*}

By using a KBBF device, a tunable VUV light source covering 5.3--7.0 eV has been recently realized in a table-top TrARPES system with an overall time resolution of 280--320 fs \cite{kbbf_bch}. Compared to UV source  generated from conventional solid-state nonlinear crystals such as $\beta$-BaB$_2$O$_4$ (BBO)  with photon energy of 5.9--6.3 eV \cite{bbo_perfetti,smallwood2012ultrafast,boschini2014innovative,yang2019time,gauthier2020tuning,bbo_bch}, the KBBF-based light source shows a larger photon energy range extending to 7.0 eV and a larger accessible in-plane momentum range \citep{zxrev2003,ARPESPrimers}. In addition, it is also more user-friendly in focusing, polarization control, although it covers a smaller accessible momentum range due to the lower photon energy compared to high harmonic generation (HHG) source generated from a noble gas  \cite{HHGWeint,EICH2014231,rohde2016time,wallauer2016intervalley,Buss2019,sie2019time,puppin2019time,Mills2019,HHGChini} and third harmonic generation (THG) source  from Xe gas \cite{cilento2016advancing,THGGEDIK} as summarized in Table \uppercase\expandafter{\romannumeral1}. Thanks to the unique widely tunable probe photon energy of the KBBF-based TrARPES system, the ultrafast dynamics of 3D Dirac fermions in Cd$_3$As$_2$ have been successfully revealed with both energy- and momentum-resolved information for the first time, and a long-lived population inversion with a lifetime of 3 picoseconds (ps) has been reported \cite{ca_bch}, demonstrating its power in revealing the electronic dynamics of 3D quantum materials.   
The overall  time resolution of the KBBF-based TrARPES system is optimized to be 280--320 fs  \cite{kbbf_bch}, which  is a major improvement for KBBF-based TrARPES system. However, such time resolution is still not ideal when compared to sub-100 fs time resolution obtained for BBO-based TrARPES systems \cite{bbo_perfetti,gauthier2020tuning,bbo_bch}, which however cannnot provide the widely tunable photon energy range as KBBF-based TrARPES system.

Achieving a time resolution better than 100 fs  while maintaining the advantage of a widely tunable probe photon energy is  important for further improving the performance of the KBBF-based TrARPES system, since the 100 fs time resolution is critical for resolving the ultrafast carrier dynamics \cite{lanzararev,zxrev} in the intrinsic timescales of the relaxation mechanisms involving electron-electron scattering \cite{rohwer2011collapse,gierz2015tracking}, electron-phonon scattering \cite{hellmann2012time,na2019direct}, and for resolving oscillations induced by coherent phonons  \cite{perfetti2006time,tbte3,fese_zx}. In this work, we report the instrumentation developments to achieve sub-100 fs time resolution in KBBF-based TrARPES system by designing a new KBBF device for fs applications. Using this newly-designed device together with systematic diagnostics and compression for the pulse duration, a VUV probe pulse with a pulse duration of 55 fs is obtained. Moreover, an overall TrARPES time resolution of 81--95 fs is achieved for the entire tunable probe photon energy of 5.3--7.0 eV, thereby extending fs KBBF device to high-performance pump-probe techniques such as TrARPES with the advantages of both high time resolution and tunable VUV laser source.

\begin{table*}[htbp]
\begin{tabular}{|c|c|c|c|c|c|}
\hline 
 \multirow{3}{*}{\diagbox{Table-top light sources}{Parameters}}&&&&&\\
   & Probe photon energy & Time resolution   & Energy resolution  & Photon flux  & Maximum accessible  \\
 & (eV)&(fs)&(meV)&(ph/s)&  in-plane momentum (\AA$^{-1}$) \\
\hline
 KBBF (This work)  &5.3--7.0 (continuous) &81--95 &35--67$^*$ &10$^{12}$--10$^{14}$ & 0.57\\[2pt] 
 BBO  \cite{bbo_perfetti,smallwood2012ultrafast,boschini2014innovative,gauthier2020tuning,bbo_bch} &5.9, 6.0, 6.2, 6.3 &58--350&11--70 &>10$^{13}$& 0.49 \\[2pt]  
 HHG  \cite{HHGWeint,EICH2014231,rohde2016time,wallauer2016intervalley,Buss2019,sie2019time,puppin2019time,Mills2019,HHGChini}   &22--42 (discrete) &13--320 &22--170 &10$^{7}$--10$^{11}$ & 2.22\\[2pt]  
 THG \cite{cilento2016advancing,THGGEDIK} &9.3, 11 &250 &16--100 &10$^{8}$--10$^{10}$& 0.92 \\[2pt] \hline
\end{tabular} 
\caption{Comparison of the performance of different table-top light sources in TrARPES system. The maximum in-plane momentum is calculated  using work function 4.5 eV and emission angle of 45$^{\circ}$.
*Note that here the energy resolution is compromised by the short pulses. A much better energy resolution can be achieved if the time resolution is sacrificed to some extent. }
\end{table*}

\section{Analysis of main contributing factors in the time resolution}

Figure 1(a) shows a schematic overview of our TrARPES system.  The fundamental beam (FB) generated from a laser oscillator with a tunable wavelength of 710 to 940 nm is frequency doubled by a BBO crystal to obtain second harmonic (SH) beam, which is frequency doubled again by a KBBF crystal to generate forth harmonic (FH) beam with a tunable photon energy of 5.3--7.0 eV, and the time resolution is compressed from $\sim$1 ps to 280 fs  \cite{kbbf_bch}. Here we focus on new improvements to further push the time resolution to below 100 fs.

The time resolution of TrARPES is determined by the duration of the pump $\Delta t_{pump}$ and probe $\Delta t_{probe}$ pulses by
\begin{equation}
\Delta t_{total} = \sqrt{(\Delta t_{pump})^2+(\Delta t_{probe})^2}.
\end{equation}
Here, the pump beam is obtained by passing the FB through a few optical components. Assuming that the chirp induced by these optical components is negligible, which is reasonable due to the relatively long wavelength, the pulse duration of the pump pulse can be approximated by $\Delta t_{pump} \approx \Delta t_{FB}$.  
The pulse duration of the probe pulse generated by KBBF device is affected by the pulse duration of the FB, and is severely broadened  by the KBBF crystal due to the large group velocity mismatch (GVM), and chirps induced by other transmissive optical components.  The above analysis suggests that three main contributing factors for the overall TrARPES time resolution are the KBBF crystal thickness which affects the probe pulse duration, the pulse duration of the FB beam which affects both the pump and probe duration, and chirps induced by the optical components which mainly affect the probe pulse duration. The effects of these three contributing factors are revealed by calculations shown in Table \uppercase\expandafter{\romannumeral2}. In the following, we further analyze these three important contributing factors, and discuss technical pathways to achieve sub-100 fs time resolution.

\begin{table}[htbp]

\centering{(a) Effect of the KBBF crystal thickness on the probe pulse duration}
\begin{ruledtabular}
\begin{tabular}{ccccc}
FB & SH  & \textbf{KBBF} & $\Delta t_{FH1}$ &\\
(fs)   &   (fs)  &\textbf{thickness} (mm) & (fs) &              \\
\hline
   &    &\textbf{1.00}   &    540        &\\
   &    &\textbf{0.50}   &    255        &\\
40 & 65 &\textbf{0.20}   &    88        &\\
   &    &\textbf{0.10}   &     54        &\\
   &    &\textbf{0.01}  &     46        &\\   
\end{tabular} 
\end{ruledtabular}

\vspace{0.5mm}
\centering{(b) Effect of the FB  pulse duration on the time resolution}
\begin{ruledtabular}

\begin{tabular}{ccccc}
KBBF    & \textbf{FB}  & SH  &  $\Delta t_{FH1}$  & Time resolution\\
thickness (mm) &    (fs)         & (fs)  &    (fs)          &     (fs)     \\
\hline
     &\textbf{40}&65   &  49 &  63    \\
0.06 &\textbf{60}&65   &  49 &  77    \\
     &\textbf{80}&73   &  54 &  97    \\
\hline
\hline

\multicolumn{5}{c}{(c) Effect of the chirp on the time resolution}\\
\hline
\hline
KBBF       & \textbf{FB} & Chirped SH  & Chirped FH   & Time resolution\\
thickness  (mm) &    (fs)         &  $\Delta t_{SH2}$  (fs) & $\Delta t_{FH2}$  (fs)    &  (fs)         \\
\hline
     &\textbf{40}&77   &  165 &  170    \\
0.06 &\textbf{60}&77   &  165 &  176    \\
     &\textbf{80}&82   &  157 &  176    \\
\end{tabular} 
\end{ruledtabular}

\caption{Calculated pulse duration and time resolution by varying KBBF thickness, FB pulse duration, and chirp compensation. (a) Effect of the KBBF thickness on FH pulse duration. The FB with a wavelength of 850 nm and a pulse duration of 40 fs is used, and the BBO thickness is 0.5 mm. The nonlinear process is calculated using SNLO software \cite{snlo}.  (b) Effect of FB pulse duration on the time resolution with a 0.06 mm KBBF, assuming that the chirps in SH and FH are fully compensated. (c) Time  resolution under the same condition as in (b) yet without compensating for the chirps in SH and FH. Here the prism length L is set to 7.5 mm.  The numbers in bold are the input parameters.}
\end{table}

 We first perform a simulation of the pulse profile and an analysis of the GVM to evaluate the effect of the KBBF crystal thickness on the probe pulse duration. The GVM induced by the KBBF crystal  during the fourth harmonic generation (FHG) is:
$$GVM = \frac{1}{v_g(FH)}-\frac{1}{v_g(SH)},$$ where $v_g(SH)$ and $v_g(FH)$ are the group velocities of the SH and the generated FH beam respectively \cite{liu2022recent}.  The positive GVM means that the FH beam falls behind the SH beam while the SH keeps generating new FH beam, leading to accumulated elongation of the FH beam.  In the extreme limit of a very thick  (1 mm) crystal, the pulse duration of the FH beam scales with the thickness of the crystal  $d$ by \cite{gauthier2020tuning}:
\begin{equation}
\Delta t_{FH1} = d_{eff}\times GVM
\end{equation}
 where $d_{eff} = d/\cos{\theta}$ and $d$ is the crystal thickness, and $\theta$ is the phase matching angle. 

Figure 2(b) shows the simulated profile of the FH beam in the time domain with different KBBF crystal thickness, where the input FB pulse is set to 40 fs with a corresponding SH pulse duration of 65 fs (frequency doubled by a 0.5 mm thick BBO), see also Table \uppercase\expandafter{\romannumeral2}(a).  For KBBF crystal thickness of 0.5 mm and 1.00 mm, the FH pulse is severely elongated and the pulse profile becomes rectangular shaped as schematically illustrated in the upper panel of Fig.~2(a). In the other limit of a thin crystal, for example, for a  KBBF thickness of 0.01 mm and a SH pulse duration of 65 fs, the frequency doubling process could lead to a shorter pulse duration than the input pulse as schematically illustrated in the bottom panel of Fig.~2(a) \cite{gauthier2020tuning}.  Figure 2(c) shows the calculated pulse duration of the probe pulse with different KBBF crystal thickness.
The pulse duration scales with the crystal thickness at thick crystal limit and reduces for thinner KBBF crystal. In order to achieve a time resolution of better than 100 fs, the KBBF thickness needs to be smaller than 0.20 mm [see also  Table \uppercase\expandafter{\romannumeral2}(a)].  For FB wavelength of 850 nm (corresponding probe energy of 5.8 eV), there is no significant improvement in the pulse duration when further reducing the crystal thickness to below 0.1 mm, while the FHG efficiency decreases for thinner crystal. However, considering that the pulse broadening from GVM becomes more significant for shorter wavelength, it is useful to choose a thinner crystal to ensure that a similar FH pulse duration can also be achieved at shorter wavelength. Taking into account all these simulation results, a KBBF crystal thickness of 0.06 mm is reasonable to achieve a time resolution of 54--98 fs in the entire probe photon energy range of 5.3--7.0 eV.

\begin{figure*}[htbp]
	\centering
	\includegraphics[] {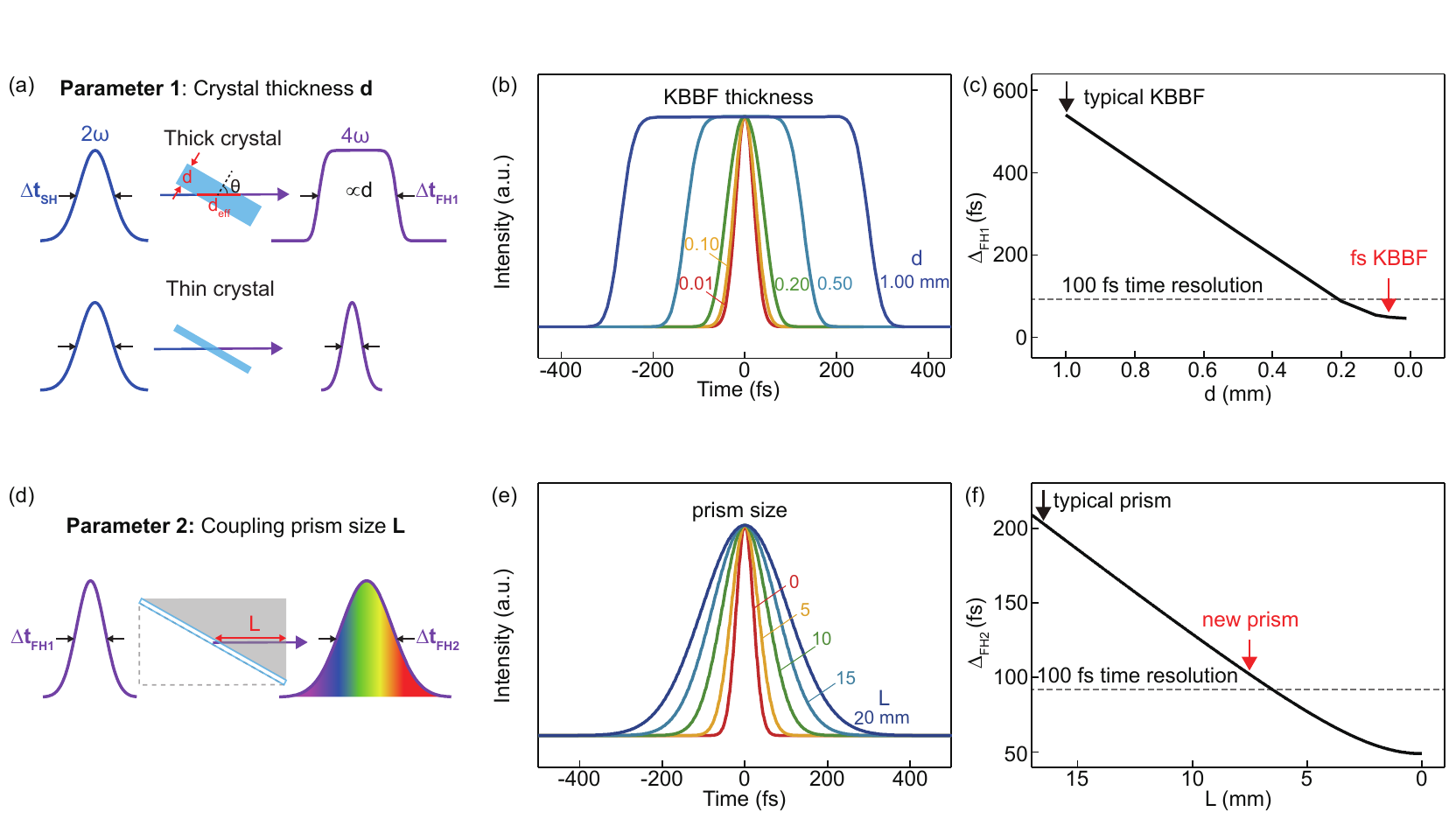}
	\label{Fig2}
	\caption{Two key parameters of KBBF devices for high time resolution. (a) Parameter 1: crystal thickness. Schematics of FHG process through thick crystal and thin crystals. (b) Calculated FH pulse profile using KBBF with different  crystal thickness [SH is set to 65 fs as shown in Table \uppercase\expandafter{\romannumeral2}(a)]. (c) Time resolution as a function of KBBF crystal thickness. (d) Parameter 2: coupling prism size.  Schematics of pulse elongation through the coupling prism. (e) Calculated FH pulse profile with different  coupling prism sizes. (f) FH pulse duration as a function of coupling prism size. The FB wavelength is set to 850 nm as an example.}
	\end{figure*} 
	
Secondly, we evaluate the effect of the FB pulse duration on the pulse duration of the probe pulse as well as the overall time resolution.  Table \uppercase\expandafter{\romannumeral2}(b)
 shows the calculated overall time resolution for KBBF thickness of 0.06 mm  at FB wavelength of 850 nm with different FB pulse duration as input parameters.  The simulation in Table  \uppercase\expandafter{\romannumeral2}(b) shows that to achieve a sub-100 fs time resolution, a pulse duration shorter than 80 fs is required for the FB. Here, we choose a FB pulse duration of 60 fs, which is the shortest possible pulse duration from this laser oscillator.

In addition to the pulse duration of the pump and probe pulses, the chirp compensation also plays a key role in the time resolution, which is illustrated by simulating the time resolution for both cases with and without adding the SH and FH compressors in Table II(b) and (c) respectively. 
The chirp corresponds to the temporal separation between different wavelengths when the beam propagates inside transmissive materials, and the elongated  pulse duration can be calculated from the group velocity dispersion (GVD) by  \cite{FHGVM}
\begin{equation}
\Delta t_{FH2} = \Delta t_{FH1} \sqrt{1+(\frac{4 ~ln2 \times L\times GVD}{\Delta t_{FH1}^2})^{^2}}
\end{equation}
where $L$ is the propagating distance and $\Delta t_{FH1}$ is the pulse duration of the generated FH pulse without chirp.  A comparison of calculation results in  Table II(b) and (c) shows that  if the chirps of SH and FH beams can be fully compensated, this can lead to an improvement of time resolution from  170 fs to 63 fs, suggesting that the chirp compensation is critical for achieving sub-100 fs time resolution. For the probe pulse, chirps can be induced by the coupling prism [P2 in Fig.~1(a)], two lenses (L2, L3), and a window in the ultra-high vacuum chamber.  The compensation for the lenses and the window has been discussed previously \cite{bbo_bch}, and here we analyze the chirp induced by the new element, mainly the coupling prism. The coupling prisms with a similar refractive index to the KBBF crystal are used before and after the KBBF crystal [Fig.~2(d)] to ensure the phase matching  condition \cite{chen2002second}, because KBBF crystals cannot be easily cut along any specific direction due to the small crystal size and the layered crystal structure. Although the chirp can  in principle be compensated by the compressor \cite{gdd}, such compensation is unlikely perfect in real experiments. Therefore, reducing the size of the coupling prism as much as possible is useful for achieving the high time resolution. Figure 2(e,f) shows the simulated FH profile and pulse duration with different coupling prism sizes (here only the prism behind FHG is considered for the pulse elongation). The spatial chirp induced by the coupling prism can be ignored because of the small divergence angle (see more details in the supplementary material).  Calculation results from Fig.~2(f) show that a large coupling prism can significantly broaden the FH pulse.  Considering the finite beam size and the space required for practical operation for photon energy tuning, we choose a coupling prism with an effective length of 7.5 mm as compared to 16.5 mm used in typical KBBF devices, thus reducing the chirp from the coupling prism significantly. Such reduction in the prism size improves FH pulse duration from 203 fs to 102 fs [a comparison between red and black arrows in Fig.~2(f)], which can be further compressed by using prism pairs to achieve sub-100 fs time resolution.

To summarize this section, the above simulation and analysis suggests that in addition to a short FB pulse duration of  $\sim$60 fs, two major improvements are critical to push  the time resolution down to below 100 fs. Firstly, a specially optimized KBBF device is required, which has a thickness of 0.06 mm and coupling prism length of 7.5 mm. Secondly, systematic diagnostics and compression are needed to further compensate for the chirp.

\section{Experimental characterization of laser beams}

\begin{figure*}[htbp]
	\centering
	\includegraphics[] {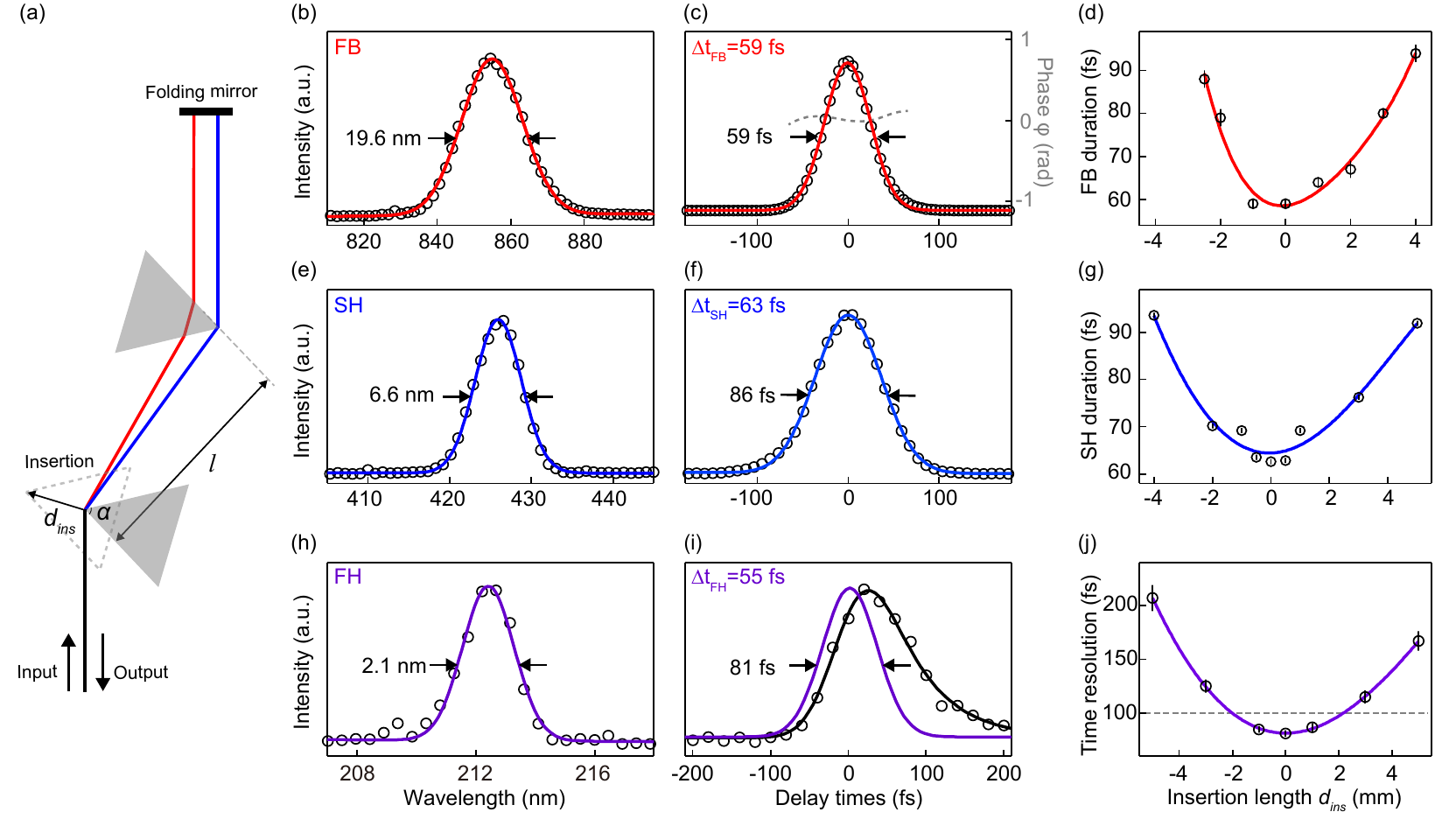}
	\caption{Pulse duration compression for FB, SH and FH. (a) A schematic of prism pair compressor. (b) Spectrum of FB. (c) Intensity and phase of FB retrieved from FROG measurement. (d) FB pulse duration as a function of prism insertion of FB compressor. The red curve is a guideline.  (e) Spectrum of the SH. (f) Cross-correlation between SH and FB from DFG measurement. (g) SH pulse duration as a function of prism insertion of SH compressor. The blue curve is a guideline.  (h) Spectrum of the FH. (i) TrARPES intensity on Sb$_2$Te$_3$ as a function of delay times by integrating the regions above 1.45 eV with pump fluence of 20 $\mu$J/cm$^2$. The purple curve is the fitted Gaussian function for extracting the time resolution. (j) Time resolution as a function of prism insertion of FH compressor. The purple curve is a guideline.  The FB is  855 nm.}
\end{figure*}

In our experimental setup, the positive chirp of the laser beam is compensated by prism pair compressors, which introduce a negative group delay dispersion (GDD)  \cite{gdd},
\begin{equation}
GDD_{prism} = \frac{4\lambda^3}{\pi c^2}(-(\frac{dn}{d\lambda})^2l+\frac{d^2n}{d\lambda^2}D)+4d_{ins}tan\frac{\alpha}{2}\times GVD_{CaF_2}
\end{equation}
where $l$ is the separation between the prism pair, $D$ is the diameter of the beam, $d_{ins}$ is the insertion length of the prism and $\alpha$ is the apex angle of the prism,  as shown in the schematic in Fig.~3(a). Here red beams with a longer wavelength travel longer in the second prism than blue beams with a shorter wavelength, adding a negative GDD to compensate for the positive chirp. The chirp compensation is obtained by adjusting the separation distance between the prism pairs $l$ and the insertion length $d_{ins}$ of the prism compressor.

Three prism pair compressors are used to compress the pulse duration of the FB, SH and FH beams respectively, and their effects are shown in Fig.~3(d,g,j). The spectrum of the FB beam shows a bandwidth of 19.6 $\pm$ 0.1 nm in Fig.~3(b), corresponding to a Fourier transform-limited pulse duration of 55 fs. The pulse duration is optimized by carefully adjusting the insertion length of the prism pair $d_{ins}$ of the FB compressor [Fig.~3(d)], and the minimum pulse duration of 59 $\pm$ 1 fs [Fig.~3(c)] is very close to the Fourier transform limit.  In addition, the retrieved phase measured by frequency-resolved optical grating [FROG, see the grey curve in Fig.~3(c)] shows negligible dependence on the delay time, confirming that there is negligible chirp after compression. 
For the SH beam, a bandwidth of 6.6 $\pm$ 0.1 nm in Fig.~3(e) corresponds to a Fourier transform-limited pulse duration of 40 fs, and the SH pulse duration is reduced from more than 90 fs to 63  $\pm$ 1 fs [Fig.~3(f,g)], showing that even though the compression is not perfect, it is still quite effective. 

Because the GVD of most optical components is much larger for shorter wavelength, for example, the GVD of CaF$_2$ is 25, 62 and 211 fs$^2$/mm for FB, SH and FH beams respectively, the chirp of FH pulse becomes very significant and requires a very large compensation. 
In the following, we use the FH as an example to illustrate how to set up the compressors. 
After FHG, the FH beam goes through a 7.5-mm thick CaF$_2$ prism, two 3-mm thick CaF$_2$ lenses and a 2.5-mm thick LiF window.  The accumulated GDD  
is 3180 fs$^2$, which determines that the prism separation $l$ should be larger than 120 mm by using  equation (4). The insertion length $d_{ins}$ can then be further used to fine tune the chirp compensation. The pulse duration of the FH beam is extracted from the TrARPES time resolution, which is measured by the temporal evolution of the high-energy TrARPES signal on Sb$_2$Te$_3$ as shown in Fig.~3(i) (see more details in the next section). The best overall time resolution reaches 81  $\pm$ 4 fs as shown in Fig.~3(i,j), which gives a FH pulse duration of 55 fs after subtracting the contribution by the FB pump beam.  The 55 fs pulse duration for the FH is similar to the previous record of 56 fs, which is achieved by broadband frequency doubling via more complicated grating pairs \cite{zhou2012generation}. Here by using a thinner crystal (0.06 mm as compared to 0.23 mm) together with careful pulse compression for the FB, SH and FH beams, we obtained the short probe pulse and implemented it for high-performance TrARPES  measurements with sub-100 fs time resolution.

\section{Sub-100 fs time resolution with tunable photon energy}

\begin{figure*}[htbp]
	\centering
	\includegraphics[]{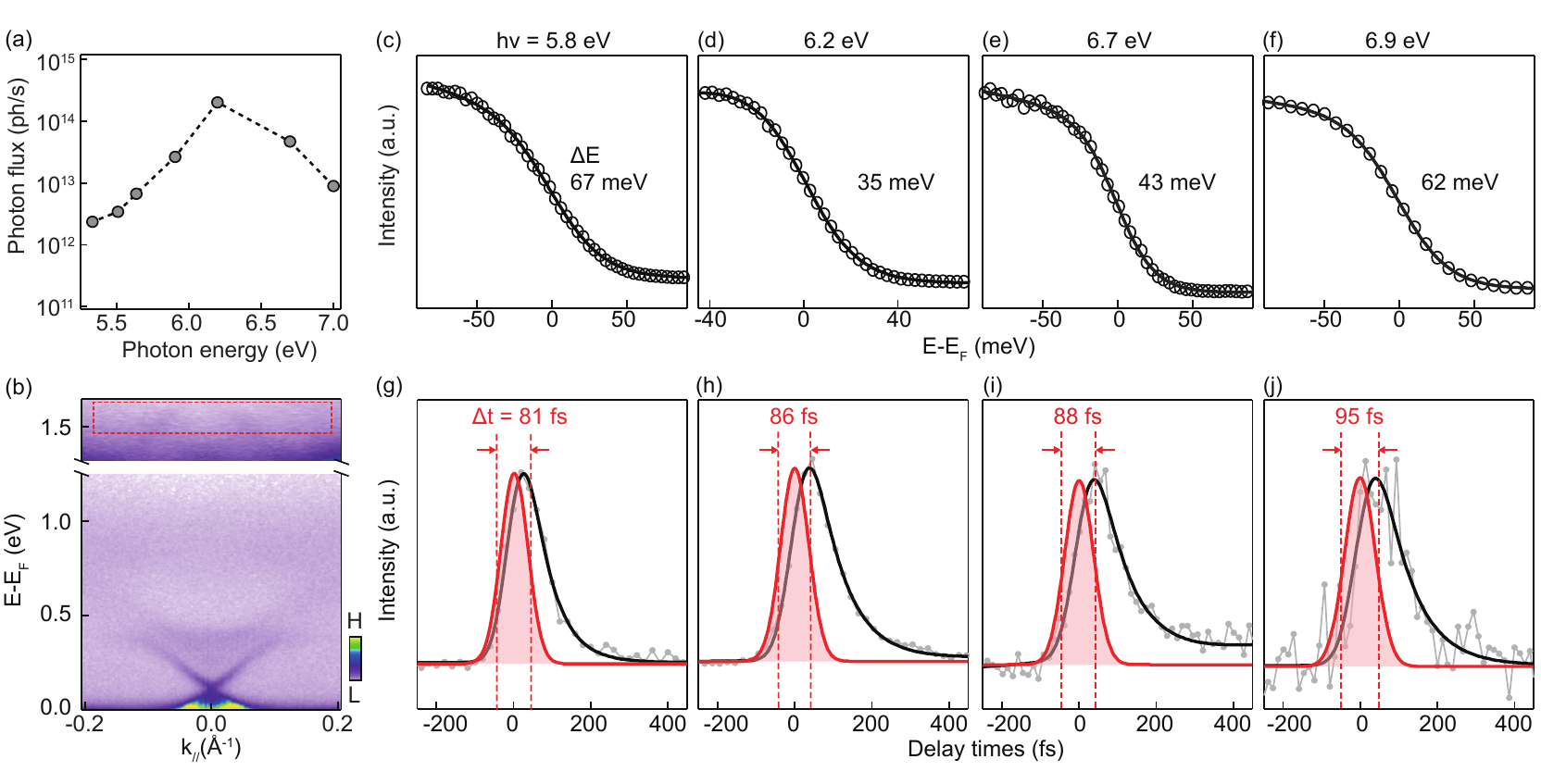}
	\caption{Tunable probe photon energy and sub-100 fs time resolution. (a) Photon flux of FH measured by power meter at 76 MHz. (b) Transient electronic structure with pump at time zero of Sb$_2$Te$_3$ with pump fluence of 80 $\mu$J/cm$^2$ and probe photon energy 6.4 eV.   (c)--(f) Energy resolutions extracted from the Fermi edge on Sb$_2$Te$_3$ at 80 K. (g)--(j) TrARPES intensity on Sb$_2$Te$_3$ as a function of delay times by integrating the regions near direct excitation ($\sim$ 1.45 eV above E$_F$) with pump fluence of 20 $\mu$J/cm$^2$. Red curves are the fitted Gaussian function for extracting the time resolution. The excitation is weak at 6.9 eV probe photon energy, which is likely due to the lack of direct excitation as discussed in the supplementary material.}
\end{figure*}

The high performance of the entire TrARPES system with the newly-designed fs KBBF device is shown in Fig.~4. Although the reduction of KBBF thickness unavoidably leads to a lower FH efficiency, the flux of the FH beam is still sufficient for TrARPES measurements  by focusing the SH beam to a diameter of 20 $\mu$m.  Figure 4(a) shows that the photon flux for 5.3--7.0 eV ranges from 10$^{12}$ to 10$^{14}$ photons/s at 76 MHz, which is high enough for high-efficiency TrARPES measurements. The energy resolution and the time resolution in the full photon energy range are extracted from measurements on a topological insulator Sb$_2$Te$_3$ film [Fig.~4(b)]. First, the energy resolution is extracted from the fitting of the Fermi edge with Fermi-Dirac distribution function. Figure 4(c)--4(f) shows the extracted energy resolution after subtracting the thermal broadening of 28 meV at the measurement temperature of 80 K. The energy resolutions of different probe photon energies range from 35 to 67 meV.
Second, the time resolution is extracted from the rising edge of TrARPES intensity evolution at the high-energy region with direct photoexcitation, whose response time is limited by the instrumentation time resolution (see more information in the supplementary material). Figure 4(g)--4(j) shows the high-energy TrARPES intensity by integrating over the red dashed box in Fig. 4(b) as a function of delay time. The time traces are fitted by the product of the Heaviside function and a single-exponential function convolved with a Gaussian function:
\begin{equation}
	I(t) = A(1+erf(2\sqrt{ln2}\frac{t-t_0}{\Delta t}-\frac{\Delta t}{4\sqrt{ln2}\tau}))e^{-\frac{t-t_0}{\tau}}+B.
\end{equation}
 to extract the time resolution \cite{bbo_bch},  where $\Delta t$ is the time resolution, $\tau$ is the relaxation time of the photo-excited electrons, $t_0$ is time zero when the pump and probe pulses overlap, A is the amplitude and B is the background. The time resolutions are extracted to be 81 $\pm$ 4, 86 $\pm$ 2, 88 $\pm$ 7 and 95 $\pm$ 15 fs for probe photon energies of 5.8, 6.2, 6.7 and 6.9 eV as shown in Fig.~4(g)--4(j), respectively. Therefore, sub-100 fs time resolution is achieved by combining the newly designed KBBF device and full diagnostics and compression for the FB, SH and FH pulses.

Finally, to check whether the resolutions are close to the Fourier transform limit, the time-bandwidth product (TBP) $\Delta$E$\cdot\Delta$t is calculated, where $\Delta$E is the bandwidth of the pulse (which is approximated by the TrARPES energy resolution here) and  $\Delta$t is the pulse duration of FH. The calculated TBP varies from 1841 to 4650 meV$\cdot$fs. The minimum value of 1841 meV$\cdot$fs at 6.2 eV is close to the Fourier transform limit of 1825 meV$\cdot$fs (see more details in the supplementary material), which allows TrARPES experiments which both high time resolution and energy resolution.

\section{Conclusions}
To summarize, by using a newly-designed fs KBBF device with a thinner KBBF crystal and smaller coupling prisms, together with full diagnostics and compression, a record high time resolution better than 100 fs is achieved in a KBBF-based TrARPES system while covering the large photon energy from 5.3 to 7.0 eV. Our work extends the application of KBBF crystal to high time resolution TrARPES measurements with tunable VUV probe source, with performance exceeding the conventional BBO-based TrARPES, thus opening up new opportunities for investigating the ultrafast dynamics of 3D quantum materials. Such fs KBBF device can also be extended to other  applications  \cite{peng2018duv} where ultrafast VUV source is important, such as time-resolved photoluminescence spectroscopy (PL) \cite{li2017enhancing}, time-resolved photoemission electron microscopy (PEEM)  \cite{KBBF_PEEM1}, photoassisted scanning tunnelling microscopy (STM) \cite{STM2}, time-resolved Raman spectroscopy  \cite{jin2014note,kash1985subpicosecond} with ultimate time resolution. 

\section*{supplementary material}
See supplementary material for the spatial chirp induced by the coupling prism of KBBF device, extraction of time resolution, the absence of direct excitation for 6.9 eV, and time-bandwidth product.
\section*{Author contribution}
Haoyuan Zhong and Changhua Bao contributed equally to this work.
\begin{acknowledgments}
This work was supported by the National Natural Science Foundation of China (Grant No.~11427903 and 11725418), the National Key R$\&$D Program of China (Grant Nos.~2021YFA1400100 and 2020YFA0308800). Changhua Bao is supported by the Shuimu Tsinghua Scholar Program.

\end{acknowledgments}

\textbf{Conflict of Interest:} The authors have no conflicts to disclose.

\textbf{Data Availability Statement:} The data that support the findings of this study are available within the article and its supplementary material.

%
\end{document}